\documentclass[sigconf]{acmart}
\def\BibTeX{{\rm B\kern-.05em{\sc i\kern-.025em b}\kern-.08emT\kern-.1667em\lower.7ex\hbox{E}\kern-.125emX}}

\usepackage{todonotes}
\usepackage{color}
\usepackage[vlined, ruled, linesnumbered]{algorithm2e}
\SetKwInOut{Parameter}{Parameters}
\usepackage{multirow}
\usepackage{verbatimbox}
\usepackage{soul}

\usepackage{amssymb}
\usepackage{pifont}
\newcommand{\cmark}{\ding{51}}%
\newcommand{\xmark}{\ding{55}}%

\copyrightyear{2020}
\acmYear{2020}
\setcopyright{acmcopyright}\acmConference[WebSci '20]{12th ACM Conference on Web Science}{July 6--10, 2020}{Southampton, United Kingdom}
\acmBooktitle{12th ACM Conference on Web Science (WebSci '20), July 6--10, 2020, Southampton, United Kingdom}
\acmPrice{15.00}
\acmDOI{10.1145/3394231.3397891}
\acmISBN{978-1-4503-7989-2/20/07}
\begin{document}

\title{How India Censors the Web}
%

\author{Kushagra Singh}
\affiliation{\institution{Centre for Internet and Society}}
\email{kushagra14056@iiitd.ac.in}
\authornote{Joint first authors}

\author{Gurshabad Grover}
\affiliation{\institution{Centre for Internet and Society}}
\email{gurshabad@cis-india.org}
\authornotemark[1]

\author{Varun Bansal}
\affiliation{\institution{Centre for Internet and Society}}
\email{varun13168@iiitd.ac.in}



%

%
\begin{abstract}
One of the primary ways in which India engages in online censorship is by ordering Internet Service Providers (ISPs) operating in its jurisdiction to block access to certain websites for its users.
This paper reports the different techniques Indian ISPs are using to censor websites, and investigates whether website blocklists are consistent across ISPs.
We propose a suite of tests that prove more robust than previous work in detecting DNS and HTTP based censorship.
Our tests also discern the use of SNI inspection for blocking websites, which is previously undocumented in the Indian context.
Using information from court orders, user reports and government orders, we compile the largest known list of potentially blocked websites in India.
We pass this list to our tests and run them from connections of six different ISPs, which together serve more than 98\% of Internet users in India.
Our findings not only confirm that ISPs are using different techniques to block websites, but also demonstrate that different ISPs are not blocking the same websites.
\end{abstract}

\keywords{Internet Censorship Analysis, Internet Service Providers, India}

\maketitle

%
%


\section{Introduction}
Nation states around the world engage in web censorship using a variety of legal and technical methods \cite{Verkamp:2012, Yadav:2018, Aryan:2013, Nabi:2013a}.
India is no different in this regard: the Government of India can legally order internet service providers (ISPs) operating in its jurisdiction to block access to certain websites for its users.
This makes the situation different from jurisdictions like Iran and China, where internet censorship is largely centralised \cite{Aryan:2013, Xu:2011}.

Legal provisions in India, namely Section 69A and Section 79 of the Information Technology (IT) Act, allow the Central Government and the various courts in the country to issue website-blocking orders that ISPs are legally bound to comply with \cite{law:69a, law:79}.
The implementation of these provisions create various uncertainties in how internet users experience web censorship.

First, the regulations do not mandate ISPs to use specific filtering mechanisms. Thus, ISPs are are at liberty to employ various technical methods \cite{Yadav:2018}.

Second, website-blocking orders, especially those issued by the Government, are rarely available in the public domain. ISPs are, in fact, mandated by regulations to maintain confidentiality of certain website-blocking orders issued by the Government \cite{rule:16}.
Various attempts by researchers and advocacy organisations to obtain the complete list of blocked websites have failed \cite{Grover:2019, SLFC:2018}.

Third, the whimsy of ISPs and the Government aggravates these problems.
Despite strict net neutrality regulations in India that prohibit ISPs from arbitrarily restricting access to websites \cite{Report:netneu}, some ISPs may be doing so nonetheless \cite{iff:2019c}.
Reports also suggest that the Government has issued blocking orders and then rescinded them on the same day \cite{Reuters:2019}.

These concerns motivated us to study web censorship in India in detail.
In particular, we seek to answer two questions pertaining to how internet users in India experience web censorship: 
(i) what are the technical methods of censorship used by ISPs in India?
(ii) are all ISPs blocking the same websites?

We contribute to research in documenting web censorship in India in three distinct ways:

\textbf{Coverage of censorship mechanisms.}
Previous work has documented the use of DNS and HTTP based censorship techniques by Indian ISPs \cite{Yadav:2018}. We include tests to determine whether ISPs are blocking websites based on the Server Name Indication, a Transport Layer Security extension.
We find that Jio (an ISP serving 49.7\% of internet users in India \cite{traiPIR:2019}) employs this technique. 
Additionally, we identify certain ISPs employing multiple censorship schemes, a fact previously undocumented in India.

\textbf{Inconsistencies in websites being blocked.}
Although previous work records some inconsistencies in website blocklists across ISPs \cite{Yadav:2018}, they use a relatively smaller corpus of potentially blocked websites (1200).
We curate the largest-known list of potentially blocked websites in India \footnote{ This list can be accessed at https://bit.ly/blockedwebsitelist} (4379), which has allowed us to draw extensive conclusions about how experiences of web censorship vary across ISPs.
We also report cases of an ISP blocking websites that are not blocked by any other ISPs. 

\textbf{Accuracy of censorship detection techniques}. Yadav et al. \cite{Yadav:2018} point out some drawbacks of relying on OONI \cite{Filasto2012a} for measuring internet censorship in India, and propose new methods of detecting DNS and HTTP based censorship.
However, we find that their methodology to detect DNS censorship makes unstated assumptions which we highlight and work around in our paper.
Additionally, their HTTP censorship detection technique relies heavily on manual inspection, making a study at our scale untenable.
We propose a novel HTTP censorship detection algorithm that requires no manual inspection, and is more accurate (in terms of F1 score) than previous automated methods.


\section{Related Work}

There have been a fair number of previous studies which explore censorship mechanisms in different countries such as China \cite{Lowe2007a, Park:2010, Xu:2011, Chai2019a}, Pakistan \cite{Nabi:2013a, Khattak:2014, Aceto2015a}, Syria \cite{Chaabane:2014}, Italy \cite{Aceto2015a}, Iran \cite{Aryan:2013}, and Korea \cite{Aceto2015a}.
Additionally, web censorship monitoring tools such as CensMon \cite{censmon:2011}, and OONI \cite{Filasto2012a} have allowed a similar analysis on a global scale.
Such works highlight that countries across the world adopt a melange of techniques to censor the web.

Amongst the most prevalent is DNS based blocking, where the network responds to DNS queries for websites it wishes to block  with either (i) DNS errors \cite{Nabi:2013a, Kuhrer:2015, Pearce:2017} or (ii) incorrect IP addresses \cite{Aryan:2013, Verkamp:2012, Wander:2014, CollateralDNS, Yadav:2018}.
Another popular technique employed by networks to filter websites is examining HTTP traffic and looking for (i) HTTP headers for blacklisted hostnames, or (ii) the HTTP request and/or response bodies for certain keywords \cite{Xu:2011, Dalek:2013a, Nabi:2013a, Aryan:2013, Khattak:2014, censmon:2011, Park:2010}.
Upon detecting such requests, censors have been found to either explicitly serve censorship notices \cite{Nabi:2013a, Aryan:2013}, close established HTTP connections \cite{Park:2010}, or both \cite{Yadav:2018}.
Some instances of filtering traffic by inspecting TCP/IP packets for destination blacklisted IP addresses have also been reported in Syria \cite{Chaabane:2014}, Italy \cite{Aceto2015a}, and China \cite{Chai2019a}.
There is also recent evidence that China is inspecting and filtering HTTPS web traffic based on the Server Name Indication present in the TLS handshake \cite{Chai2019a}, but previous work has reported that no Indian ISP uses this technique \cite{Yadav:2018}.

In the Indian context, there has been an initial attempt to understand the censorship mechanisms employed by Indian ISPs \cite{Yadav:2018, Gosain:2017}.
However, as we find, these studies have not uncovered the full extent of web censorship in India in terms of both technical mechanisms and scale. 

\section{Data Curation}
\label{sec:data}
We compile a list of of potentially blocked websites from three types of sources:

\textbf{Government orders.} A website/URL blocking order may come from the Government of India \cite{law:69a, law:79}.
These orders are usually not in the public domain.
For orders issued under section 69A of the IT Act, a confidentiality clause prevents any party from disclosing its contents \cite{rule:16}.
We collect published and leaked Government orders that are available publicly, which contribute 890 URLs to our corpus.

\textbf{Court orders}. The various courts in India also have the power to issue website blocking orders\cite{law:79}.
Not all such orders are available in the public domain \cite{iff:2019b}.
However, the Government and BSNL (a public company operating as an ISP) have provided portions of this list when under pressure to respond to Right to Information (RTI) requests. \cite{SLFC:2018, iff:2019a}.
Court orders contribute 9367 URLs to our list.

\textbf{User reports.} The Internet Freedom Foundation collects and publishes reports from internet users who notice blocked websites\footnote{ https://bit.ly/iffuserreport}.
These contribute an additional 62 URLs to our list.

Collecting data from these sources led to a total of 9673 unique URLs.
Given that most of these URLs are sourced from recent court orders, there is a high possibility of them being currently blocked.
The scope of our analysis is restricted to website-level (rather than webpage-level) blocking, and so we extract unique domain names from this list, resulting in 5798 websites.
To limit ourselves to active websites, we exclude all websites for which we could not resolve via Tor circuits. 
We end up with a total of 4379 websites, which to the best of our knowledge, is the largest known corpus of potentially blocked websites in India. 

\section{Methodology}
We probe for the presence of different censorship techniques in six major Indian ISPs.
These include Reliance Jio Infocomm (Jio), Bharti Airtel (Airtel), Vodafone Idea (Vodafone), Bharat Sanchar Nigam Ltd. (BSNL), Atria Convergence Technologies (ACT), and Mahanagar Telephone Nigam Ltd. (MTNL).
The Telecom Regulatory Authority of India reveals that as of October 2019, these six ISPs together serve 657.46 million internet subscribers in India, i.e. 98.82\% of the subscriber base in India \cite{traiPIR:2019}.

\subsection{DNS censorship}
Domain Name System (DNS) resolution involves translating a hostname to its corresponding IP address(es), and is usually the first step in accessing a website.
Traditional DNS resolution is prone to poisoning and injection attacks \cite{rfc3833}.
There are secure resolution protocols such as DNSSEC \cite{dnssec:rfc4033}, DNS over HTTPS (DoH) \cite{doh:rfc8484}, and DNS over TLS (DoT) \cite{dnstls:rfc7858} that mitigate these attacks; however, they are not widely deployed \cite{dnssecadoption:2019, Lu:2019}.

\textbf{DNS Poisoning.} By default, DNS queries are sent to a resolver configured by the ISP.
Thus, ISPs can return an incorrect IP address or nothing at all in response to clients' DNS queries for websites they wish to block \cite{BoruDNSSpoof, Aryan:2013, Verkamp:2012}.

\textbf{DNS Injection.} ISPs can intercept DNS queries for websites they wish to block and inject incorrect IP addresses in the responses \cite{CollateralDNS, Wander:2014, Pearce:2017}.

We term an ISP's use of DNS poisoning or DNS injection attacks to block websites as DNS censorship.

\subsubsection{\textbf{Existing techniques}}
Detecting DNS censorship has previously been done by comparing responses from the test resolver with responses from trusted resolvers \cite{Filasto2012a, Aceto:2017}.
However, this can lead to an over-reporting in censorship as these trusted resolvers can respond with a different IP address for legitimate reasons (such as load balancing) \cite{Ager:2010}.
Lowe et al. circumvent this problem by selecting 5 censorship-free control resolvers and only investigating domain names for which all resolvers returned the same IP address; however, this results in a decrease in the size of the test list \cite{Lowe2007a}.

Another technique is to rely on the autonomous system (AS) number\footnote{ https://www.apnic.net/get-ip/faqs/asn/} to which the returned IP address belongs. 
Kuhrer et al. \cite{Kuhrer:2015} consider a DNS response legitimate if the IP addresses returned via the trusted and tested resolvers belong to the same AS.
This approach fails to take into account that a domain name can resolve to IP addresses belonging to different ASes.
Yadav et al. deem a DNS response censored if the returned IP address belongs to the same AS as the client's IP address \cite{Yadav:2018}. 
This approach rides on two unsubstantiated assumptions: (i) that the incorrect IP address returned by the ISP always belongs to the same AS as the client's, and (ii) the given website is not hosted within the same AS.
Our proposed technique works around all these flaws.

\subsubsection{\textbf{Proposed technique}} 
\label{section:dnsCollection}
We begin by creating a set of IP addresses $IP_{d,C}$ for each domain name $d$ in our list by combining the responses obtained by resolving it via 5 censorship-free control networks (collectively termed $C$): (i) Tor circuits with exit nodes in US, CA and AU; and (ii) DoH servers run by Cloudflare and Google.
In a test network, we attempt to resolve each domain name in our list using the ISP-assigned resolver.
If the resulting IP address is present in $IP_{d,C}$, we conclude that the ISP is not using DNS censorship to block that website.
Otherwise, similar to \cite{Kuhrer:2015, Yadav:2018}, we flag the domain name censored if (i) the resolver responds with an error (for eg. \texttt{NXDOMAIN}), or (ii) the resolver responds with a bogon IP\footnote{https://ipinfo.io/bogon}.

We further investigate the list of domain names $D'$, for which the ISP-configured resolver returns an IP address not found in $IP_{d,C}$.
For these domain names, the test network is returning either (i) a legitimate IP address not captured via the control networks, or (ii) an incorrect IP address. We term the second possibility as DNS tampering.
As \cite{Yadav:2018, Aryan:2013} report, ISPs implementing DNS tampering respond with the same incorrect IP address to DNS queries for websites it wishes to block.
We try to identify such behaviour by an ISP by looking at all IP addresses being returned by it for domain names in $D'$.

For each network $n$, we construct $IP_n$, the list of IP addresses received by resolving domain names in $D'$ via that network.
Next we calculate $MRF_n$, the relative frequency of the most frequent IP address in $IP_n$.
By comparing $MRF$ values of the test and control networks, we are able to discern if the test network responds with an abnormally recurring IP address.
This would be characteristic of an ISP that is censoring websites by returning an incorrect IP address, i.e. employing DNS tampering\footnote{ We leverage the fact that most websites in our curated list have a high possibility of being blocked. See section \ref{sec:data}}.
In ISPs where we detect DNS tampering, we mark domain names for which the DNS response was the most frequent IP address as censored.

\begin{center}
    \SetAlFnt{\small}
    \begin{algorithm}
    \label{algo:dns}
    \caption{DNS Tampering Detection}
    \DontPrintSemicolon
    \KwIn{Test Network $T$, Control Networks $C$, Domain Names $D$}
    \KwResult{Determines DNS tampering}
    \SetKwProg{Fn}{Function}{}{end}
    
    \SetKwFunction{algo}{algo}
    \SetKwProg{myalg}{Algorithm}{}{}

        $C \leftarrow \{C_1, C_2, ..., C_k\}$ \tcp*{control networks (Tor, DoH)}

        \For{$d \in D$}{
            \For{$c \in C$}{
                $IP_{d,c} \leftarrow$ DNS response for $d$ collected via $c$\ \\
            }
            $IP_{d,C} \leftarrow \{ IP_{d,c} | c \in C\}$ \\
            $IP_{d,t} \leftarrow $ DNS response for $d$ collected via test network \\
        }
        $D' \leftarrow \{d | d \in D \wedge IP_{d,t} \notin IP_{d,C} \}$ \tcp*{domain names for which test response did not match any of control responses}
        
        \For{$c \in C$}{
            \For{$d \in D'$}{
                $IP_{d,c} \leftarrow$ DNS response for $d$ collected via $c$\ \\
            }
            $IP_c \leftarrow \{IP_{d,c} | d \in D' \}$  \\
            \For{$ip \in IP_c$}{
                $RF_{ip,c} \leftarrow$ Relative frequency of $ip$ in $IP_c$ \\
            }
            $MRF_c \leftarrow max(\{RF_{ip,c} | ip \in IP_c\})$
        }
        
        $\mu_{MRF_C} \leftarrow$ mean of $MRF_c$ $\forall$ $c$ $\in$ $C$ \\
        $\sigma_{MRF_C} \leftarrow$ standard deviation of $MRF_c$ $\forall$ $c$ $\in$ $C$ \\

        \For{$d \in D'$}{
            $IP_{d,t} \leftarrow$ DNS response for $d$ in test network \\
        }
        $IP_t \leftarrow \{IP_{d,t} | d \in D' \}$ \\
        \For{$ip \in IP_t$}{
            $RF_{ip,t} \leftarrow$ Relative frequency of $ip$ in $IP_t$ \\
        }
        $MRF_t \leftarrow max(\{RF_{ip,t} | ip \in IP_t\})$ \\
        
        \uIf{$MRF_t - \mu_{MRF_C} > 3*\sigma_{MRF_C}$}{
            \Return DNS Tampering present \\
        }
        \Else{\Return DNS Tampering not present \\}
    \end{algorithm}
\end{center}

\subsection{TCP/IP Blocking}
ISPs can block access to websites by preventing clients from connecting to the specific IP addresses the website is hosted on \cite{censmon:2011, Chaabane:2014, Aceto2015a, Chai2019a}.
Additionally, the ISP may inspect TCP packet headers for the destination port number if it wishes to block certain types of traffic for that IP address. 
However, this censorship technique can result in over-blocking due to the popularity of virtual hosting, which allows multiple websites to be hosted on the same IP address \cite{Edelman:2003}.
These pitfalls are a plausible explanation for why ISPs in Korea \cite{Aceto2015a}, Iran \cite{Aryan:2013} and Pakistan \cite{Nabi:2013a, Aceto2015a} do not use this technique.
Yadav et al also conclude the same for Indian ISPs\cite{Yadav:2018}; however, it is unclear how they determine what IP addresses to probe to detect such censorship.

\subsubsection{\textbf{Proposed technique}}
We first obtain legitimate IP addresses for websites in our test list as discussed in section \ref{section:dnsCollection}.
This precludes any DNS censorship by ISPs from interfering with our test.

Building on \cite{censmon:2011, Yadav:2018}, we perform a two-step test.
First, we ping\footnote{ https://linux.die.net/man/8/ping} the IP address to verify whether it is reachable through the test network.
A response implies that the ISP is not filtering traffic based on the destination IP address.
For such IP addresses, we then attempt to establish a TCP connection on ports 80 (used for HTTP traffic) and 443 (used for HTTPS traffic).
A successful TCP 3-way handshake with a given IP address and port would imply the absence of TCP-based blocking.
A failure in either step, however, can be attributed to network congestion, host unavailability or, of course, censorship by the ISP.
To establish that connection failures are indeed due to censorship, we run the same tests via Tor circuits with exit nodes in censorship-free countries (USA, CA and AU).
We rule out host unavailability and network congestion by reattempting failed connections five times with a delay of 100 seconds.

\subsection{HTTP Filtering}
Unencrypted HTTP traffic between a client and host can be intercepted and monitored.
Previous studies have discovered different techniques adopted by censors for blocking HTTP access \cite{Nabi:2013a, Aryan:2013, Khattak:2014, Park:2010, censmon:2011}.
When a client attempts to access a website that the ISP seeks to block, the ISP sends back forged TCP or HTTP packets that seem to be originating from the host. 
These can include (i) a TCP packet with the \texttt{RST} (reset) bit set, forcing the client to kill the connection instantly \cite{Park:2010}, (ii) an HTTP 2xx response\textsuperscript{\ref{note1}} containing a censorship notice \cite{Nabi:2013a}, (iii) an HTTP 3xx response\footnote{\label{note1} HTTP response codes (RFC 7231) -- https://tools.ietf.org/html/rfc7231\#section-6.1} redirecting the client to a URL serving a censorship notice \cite{Nabi:2013a}, or (iv) an HTTP 4xx/5xx response\textsuperscript{\ref{note1}} conveying an HTTP error to the client \cite{Aryan:2013}.

\subsubsection{\textbf{Existing techniques}}
Due to the size of our corpus, we cannot rely on manual inspection as done by \cite{Yadav:2018}.
Existing automated techniques for detecting censored HTTP responses usually rely on making comparisons with uncensored responses, collected either via control servers set up in censorship-free countries \cite{censmon:2011, Aceto2015a}, or via Tor circuits \cite{Yadav:2018, Filasto2012a, Ververis:2015, Esnaashari:2013}.
Due to the dynamic nature of content hosted on websites, comparing verbatim responses can be erroneous \cite{Jones:2014b}.
Moreover, the geographical location of a client can also introduce variations (such as the content language) in the received response.
To mitigate these issues, prior research utilizes meta information derived from responses for comparison \cite{Jones:2014b, Filasto2012a, Ververis:2015}.

Jones et al. propose methods for identifying HTML pages which contain a censorship notice \cite{Jones:2014b}.
They report that comparing a test response's length and HTML DOM structure with that of an uncensored response can help identify such pages with a high accuracy.
Similarly, other authors use response length in conjunction with different HTML similarity metrics for comparisons \cite{Aceto2015a, Esnaashari:2013, Filasto2012a}.
However, these approaches perform well only in instances where censors explicitly inject censorship notices, an assumption that generally doesn't hold true: as discussed above, censors have been known to adopt tacit approaches such as responding with HTTP errors or redirections.
In the absence of HTML responses, these methods depend solely on response lengths, which as \cite{Yadav:2018} discover, can be inefficacious \cite{Yadav:2018}.
The OONI tool \cite{Filasto2012a} does a more elaborate comparison, drawing conclusions by observing differences in status codes, headers, lengths, and HTML titles.
However, even their approach culminates in false positives and false negatives \cite{Yadav:2018}.

Building upon these approaches, we propose a more robust automated technique for detecting censored responses (outlined in algorithm \ref{algo:http}) and use it to probe Indian ISPs for HTTP censorship.

\subsubsection{\textbf{Collecting HTTP responses}}
We begin by resolving the IP address of each domain name in our list via trusted resolvers, as discussed in section \ref{section:dnsCollection}.
Since DNS resolution may return multiple IP addresses for the same domain name, we probe all resulting \texttt{(domain name, IP address)} pairs in our experiment.

For each \texttt{(domain name, IP address)} pair, we make HTTP GET requests to the IP address, with the \texttt{HOST} header set as the domain name.
This is done via 5 control servers in censorship-free countries (US, CA, GB, NE and AU), and via the test network.
Unlike some studies \cite{Yadav:2018, Filasto2012a, Ververis:2015, Esnaashari:2013} we avoid using Tor circuits for collecting control responses, since some websites blacklist them and respond differently to HTTP requests originating from them.
Additionally, instead of using just one control response \cite{Yadav:2018, Filasto2012a, Jones:2014b}, we consider multiple responses.

\subsubsection{\textbf{Proposed technique}}
After collecting the control and test responses, we follow the detection technique outlined in algorithm \ref{algo:http}.
First, we compare the HTTP status code of control responses with that of the test response.
If these status codes do not match, we classify the test response as censored.
However, the opposite need not necessarily imply the absence of censorship.
In case the status codes are the same, we investigate further on a case by case basis as explained below.
\begin{itemize}
    \item \textbf{2xx (Success):} We check for response length inconsistency and response body inconsistency.
    \item \textbf{3xx (Redirection):} We compare the domain name present in the redirect URLs. 
    \item \textbf{4xx/5xx (Error):} We compare the session header keys.
\end{itemize}

\begin{center}
    \SetAlFnt{\small}
    \begin{algorithm}[h]
    \caption{HTTP Censorship Detection}
    \label{algo:http}
    \KwIn{DOMAIN NAME dn, IP ADDRESS ip}
    \KwResult{Determines HTTP censorship}
    control\_res $\leftarrow$ HTTP GET response for dn,ip in control networks\;
    test\_res $\leftarrow$ HTTP GET response for dn,ip in test network\;
    \If{connection reset while getting test\_res} {
        \Return \textit{censored}\;
    }
    \uIf{control\_res.status\_code $\neq$ test\_res.status\_code}{
        \Return \textit{censored}\;
    }
    \uElseIf {test\_res.status\_code = 2xx} {
        \uIf{test\_res.length inconsistent OR test\_res.body inconsistent}{\Return censored\;}
        \Else {\Return uncensored\;}
    }
    \uElseIf {test\_res.status\_code = 3xx} {
        \uIf{mismatch in control\_res, test\_res redirect HOSTNAMEs}{\Return \textit{censored}\;}
        \Else {\Return \textit{uncensored}\;}
    }
    \uElseIf {mismatch in control\_res, test\_res header keys} {\Return \textit{censored}\;}
    \Else {\Return \textit{uncensored}\;}
    \end{algorithm}
\end{center}


The response length inconsistency and response body inconsistency used above is defined as follows

\textbf{Response length inconsistency.} Given control response lengths ($L_{c_i},L_{c_{ii}},...L_{c_{n}}$) and a test response length $L_t$, we call $L_t$ inconsistent if $|\mu_{L_c} - L_t| > 3 * \sigma_{L_c}$.
Here $\mu_{L_c}$ is the mean, and $\sigma_{L_c}$ the standard deviation of the control response lengths.

\textbf{Response body inconsistency.} For each control and test response, we generate term frequency ($TF$) vectors using HTML tags extracted from the response body. 
A test response body is called inconsistent if $|\mu_{c,c} - \mu_{t,c}| > 3 * \sigma_{c,c}$, where $\mu_{c,c}$ is the mean cosine similarity between $TF$ vectors of control responses, $\mu_{t,c}$ the mean cosine similarity between $TF$ vectors of test and control responses, and $\sigma_{c,c}$ the standard deviation of cosine similarity between $TF$ vectors of control responses.

To verify the efficacy of our proposed technique, we manually inspect 500 responses from the six ISPs (a total of 3000 responses), and categorise them as censored or uncensored.
We implement the previous techniques and compare their predictions on the annotated set to ours.
As reported in table \ref{table:httpApproachResults}, our proposed technique detects both censored and uncensored responses with a higher f1-score than previous approaches.

\begin{table}[h]
    \resizebox{0.9\textwidth}{!}{\begin{minipage}{\textwidth}

    \begin{tabular}{|c|c|c|c|c|c|c|}
    \hline
    \multirow{2}{*}{Detection Technique}   & \multicolumn{2}{c|}{Precision}    & \multicolumn{2}{c|}{Recall}   & \multicolumn{2}{c|}{F1 score}   \\ \cline{2-7} 
        & C     & U     & C     & U     & C     & U         \\ \hline
    Length difference \cite{Jones:2014b, Yadav:2018}   & 0.65  & 0.73  & 0.77  & 0.59  & 0.70  & 0.66  \\ \hline
    HTML similarity \cite{Jones:2014b}  & 0.45  & 0.44  & 0.62  & 0.28  & 0.52  & 0.34  \\ \hline
    OONI \cite{Filasto2012a}   & 0.67  & 1.00  & 1.00  & 0.54  & 0.80  & 0.70  \\ \hline
    \textbf{Proposed}    & \textbf{0.71}  & 0.98  & 0.99  & \textbf{0.63}  & \textbf{0.83}  & \textbf{0.77}  \\ \hline
    \end{tabular}
    
    \end{minipage}}
    \caption{Performance of various HTTP censorship detection techniques. We report Precission, Recall and F1-score for Censored (C) and Uncensored (U) classes. Our proposed technique has a higher F1-score than the previous techniques.}
    \label{table:httpApproachResults}
    
\end{table}

\subsection{SNI Based Censorship}
The Server Name Indication (SNI) was designed as an extension to TLS to support the hosting of multiple HTTPS websites on the same IP address \cite{rfc6066}.
The SNI is an attribute included in the \texttt{ClientHello} message, where the client specifies the hostname it wishes to connect to.
Since the SNI is in clear-text, censors can monitor this field for hostnames and block websites by preventing successful TLS connections \cite{Shabir:2015, Chai2019a, Zolfaghari:2016, rfc8404}.
While such censorship has been documented in China \cite{Chai2019a} and South Korea \cite{Korea:2019}, there has been no prior evidence to suggest that Indian ISPs are using this technique \cite{Yadav:2018}.

\subsubsection{\textbf{Proposed technique}}
For this test, we take advantage of a server configured to accept TLS  connections even if it does not host the website specified in the SNI.
For each potentially blocked website, we attempt to establish a TLS version 1.3 connection with that server's IP address using the website's hostname as the SNI.
A successful connection would imply the absence of SNI-inspection based censorship in the test network.
Using TLS version 1.3 ensures that only the IP address and servername are present in clear-text, since all subsequent communication after the \texttt{ClientHello} and \texttt{ServerHello} is encrypted \cite{rfc8446}.
This precludes the interference of any other censorship technique used by the ISP with our test.

\section{Results}
We first report the different censorship techniques adopted by ISPs, and then compare the consistency of website blocklists across ISPs.

\subsection{Censorship techniques}
\label{section:cenTech}
We notice stark differences in censorship mechanisms adopted by Indian ISPs, each using a range of techniques individually or in combination to censor websites (outlined in Table \ref{table:ISPTechniques}).

\begin{table}[h]
\centering
\begin{tabular}{|c|c|c|c|c|}
\hline
ISP      & DNS   & TCP/IP & HTTP  & SNI   \\ \hline \hline
ACT      & \cmark & \xmark  & \cmark & \xmark \\ \hline
Airtel   & \cmark & \xmark  & \cmark & \xmark \\ \hline
BSNL     & \cmark & \xmark  & \xmark & \xmark \\ \hline
Jio      & \xmark & \xmark  & \cmark & \cmark \\ \hline
MTNL     & \cmark & \xmark  & \xmark & \xmark \\ \hline
Vodafone & \xmark & \xmark  & \cmark & \xmark \\ \hline
\end{tabular}
\caption{Censorship techniques employed by Indian ISPs}
\label{table:ISPTechniques}
\end{table}

\begin{figure}[h]
  \centering
  \includegraphics[width=0.95\linewidth]{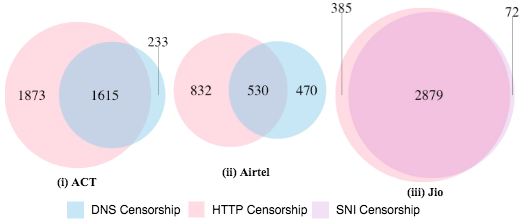}
  \caption{Censorship techniques used by (i) ACT, (ii) Airtel, and (iii) Jio for blocking websites. We notice the same ISP using multiple techniques for blocking different websites. }
  \label{fig:venn}
  \label{fig:inconsistent}
\end{figure}

Interestingly, we notice different trends as compared to the previous study on Internet censorship in India \cite{Yadav:2018}: (i) They report Airtel to be using only HTTP header inspection based censorship for blocking websites. We notice otherwise, with Airtel using DNS censorship in conjunction with the aforementioned technique. (ii) They report no instances of SNI-inspection based censorship in any ISP, whereas we observe Jio to be using it extensively for blocking (2951 websites).
These new observations indicate an evolving nature of censorship mechanisms employed by Indian ISPs.
Further, we notice that all ISPs using multiple censorship mechanisms are not blocking the same websites with each mechanism.
For instance, ACT uses only DNS censorship for blocking 233 websites, only HTTP censorship for 1873 websites, and both to block 1615 websites.
Such irregularities are illustrated in figure \ref{fig:inconsistent}.

\subsubsection{\textbf{DNS}}
\begin{figure}[b]
  \centering
  \includegraphics[width=1\linewidth]{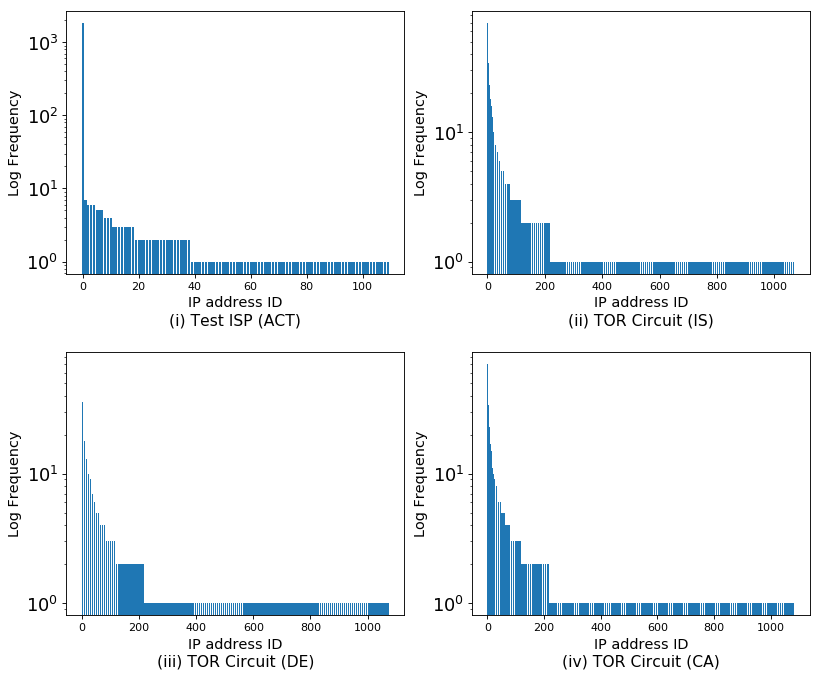}
  \caption{Log frequency plot of IP addresses received by resolving websites in our test list. We notice an abnormally large spike in subplot (i) (corresponding to ACT), compared to subplots (ii), (iii), and (iv) (corresponding to Tor circuits)}
  \label{fig:freqPlot}
  
\end{figure}

We observed DNS censorship in four ISPs: ACT, Airtel, BSNL, and MTNL.
Airtel is unique in responding with \texttt{NXDOMAIN} errors to DNS queries for websites it blocks.
In the other three ISPs (ACT, BSNL, and MTNL), we found that a distinct IP address appeared unusually frequently in DNS responses when we tried to resolve potentially blocked websites using the ISP-assigned resolver.
This was in line with our intuition as detailed in section \ref{section:dnsCollection}.
By comparing the relative frequency of the most frequently occurring IP address in responses collected from the test to those collected from the control networks, we were able to detect which ISPs were using DNS tampering.
For illustration, Figure \ref{fig:freqPlot} compares the frequency of IP addresses received for DNS queries for potentially-blocked websites in four networks: an ISP that uses DNS-based censorship (ACT), and three Tor circuits with exit nodes in censorship-free countries.
Each of these three ISPs responded with a unique incorrect IP address.
Using this fact, we conclude that there was no collateral censorship from DNS-based blocking by other ISPs.

\subsubsection{\textbf{TCP/IP}}
Given the immense collateral censorship caused by TCP/IP based methods of censorship, we were not surprised to find that, in line with \cite{Yadav:2018}, no ISP we investigated uses this technique.

\subsubsection{\textbf{HTTP}}
We observe HTTP-header based censorship in all the six ISPs we investigated, but find only ACT, Airtel, Jio, and Vodafone to be serving distinct censorship notices.
Additionally, we notice Airtel to be closing connections and not serving any censorship notice for the majority of the websites it blocks.
Using the unique signatures of these responses, we are able to identify collateral censorship in other ISPs stemming from these ISPs.
For instance, we find that all instances of HTTP censorship we detect in BSNL and MTNL are attributable to Airtel and ACT.
There was also a small number of instances where we observed Vodafone's censorship notices from tests run through Jio (2), and Airtel's notices in tests run through Vodafone (2).

\subsubsection{\textbf{SNI Inspection}}
We observe SNI inspection based censorship only in one ISP, Jio.
Since we did not observe SNI-based censorship in any another ISP, we rule out collateral censorship caused by Jio's employment of this technique.
Out of 3340 websites we found Jio to be censoring, we notice the use of SNI inspection for 2951 websites.

\subsection{Website blocklists}
If we observe a particular ISP blocking a website by any method, we mark it as censored by that ISP. 
We term this list of censored websites by a particular ISP as its website blocklist.
Note that for an ISP's blocklist, we only consider websites are censored by the ISP using its own mechanisms, i.e. we ignore collateral censorship which we highlighted in section \ref{section:cenTech}).

From our list of 4379 potentially blocked websites that we tested, we find that 4033 appear in at least one ISP's blocklist. We use this list of 4033 websites for further analysis to see whether website blocklists are consistent across ISPs.

\begin{figure}[h]
  \centering
  \includegraphics[width=0.9\linewidth]{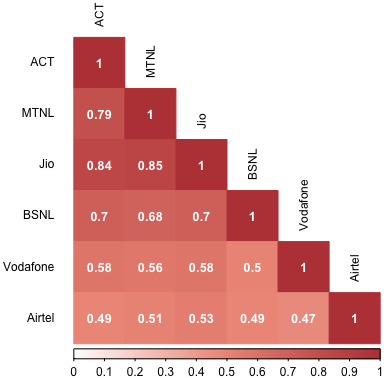}
  \caption{Heatmap illustrating the overlap of blocklists of different ISPs. For each pair of ISP blocklists $L_a$ and $L_b$, we calculate the Jaccard similarity coefficient, i.e. $ \frac{|L_a \cap L_b|}{|L_a \cup L_b|}$.}
  \label{fig:variation}
  
\end{figure}

Interestingly, we notice large inconsistencies in ISPs' blocklists. 
For instance, we find that in terms of absolute numbers, ACT blocks the maximum number of websites (3721).
Compared to ACT, Airtel blocks roughly half the number of websites (1892).
Table 2 notes the size of each ISP's blocklist.
Perhaps most surprisingly, we find that only 1115 websites out of the 4033 (just 27.64\%) are blocked by all six ISPs.
Figure \ref{fig:variation} illustrates the variation in different ISPs' blocklists.

We also found that several websites (215) are being blocked by only a single ISP out of the six. For instance, ACT blocks 62 websites that are not blocked by another ISP. This calls into question whether blocking of these websites has any standing legal basis, and is potential evidence of the fact that ISPs are blocking websites arbitrarily.

\begin{table}[h]
\begin{tabular}{|c|c|c|c|c|c|}
\hline
ACT & Airtel & BSNL & Jio & MTNL & Vodafone \\ \hline
3721 & 1892 & 3033 & 3340 & 3182 & 2273 \\ \hline
\end{tabular}
\caption{Number of websites (out of 4033) blocked by ISPs}
\label{table:blocklist}

\end{table}

\section{Conclusion}

Our work presents the largest study of web censorship in India, both in terms of number of censorship mechanisms that we test for, and the corpus size of the potentially-blocked websites.
In terms of censorship methods, our results confirm that ISPs in India are at liberty to use any technical filtering mechanism they wish: there was, in fact, no single mechanism common across ISPs.
We also found a deep packet inspection technique, namely SNI inspection, already being employed by the largest ISP in India (Jio) to censor websites.
Our work also records large inconsistencies in website blocklists across ISPs in India.

Simply stated, we find conclusive proof that Internet users in India can have wildly different experiences of web censorship.

Analysing inconsistencies in blocklists also makes it clear that ISPs in India are (i) not properly complying with website blocking (or subsequent unblocking orders), and/or (ii) arbitrarily blocking websites without the backing of a legal order. This has important legal ramifications: India's net neutrality regulations, codified in the license agreements that ISPs enter with the Government of India\cite{Report:netneu}, explicitly prohibit such behaviour. Thus, our work provides empirical evidence of the fact that Indian ISPs may be violating net neutrality regulations.

Our work also points to how the choice of technical methods used by ISPs to censor websites can decrease transparency about state-ordered censorship in India.
While some ISPs were serving censorship notices, other ISPs made no such effort.
For instance, Airtel responded to DNS queries for websites it wishes to block with NXDOMAIN.
Jio used SNI-inspection to block websites, a choice which makes it technically impossible for them to serve censorship notices.
Thus, the selection of certain technical methods by ISPs exacerbate the concerns created by the opaque legal process that allows the Government to censor websites.

Web censorship is a restriction on the right to freedom of expression and the right to access information, which are guaranteed to all citizens by the Constitution of India.
There is an urgent need to reevaluate the legal and technical mechanisms of web censorship in India to make sure the curtailment is transparent, and the actors accountable.

\section{Future Work}
Recent user reports have suggested that website blocklists may vary within the same ISP based on geographical location.
Additionally, this variance in blocklists may also occur within mobile and broadband networks belonging to the same ISP.
Future work may involve running our tests from different vantage points in the country to determine the extent of such vagaries.

Contrasting results from previous studies also seem to suggest an evolving nature of the internet censorship mechanism in India.
Such an evolving mechanism adopted by the ISPs demands that censorship-evasion techniques also adapt in tandem. Future research can focus on developing tools that help Indian netizens evade website blocking.

\begin{acks}
We thank Devashish Gosain, Tushar Kataria, Divyank Katira and three anonymous reviewers for their constructive feedback and suggestions.
We are also grateful to Suhan S and Harikarthik Ramesh for helping curate the list of potentially blocked websites used in the tests. We also thank the Internet Freedom Foundation for collecting and maintaining user reports on blocked websites, and IPinfo.io for giving us access to their IP address data.

\end{acks}
\newpage
%
\bibliographystyle{ACM-Reference-Format}
\bibliography{main}
\end{document}